\documentclass[aps,preprint,showpacs]{revtex4}
\usepackage[T1]{fontenc}
\usepackage[latin1]{inputenc}
\usepackage{graphics}
\usepackage{setspace}

\makeatletter

\providecommand{\LyX}{L\kern-.1667em\lower.25em\hbox{Y}\kern-.125emX\@}

\makeatother

\begin{document}

\title{Theoretical Study of Magnetism and Superconductivity in \protect\( 3d\protect \)
Transition -Metal-\protect\( MgB_{2}\protect \) Alloys}

\author{Prabhakar P. Singh and P. Jiji Thomas Joseph }

\affiliation{Department of Physics,  Indian Institute of Technology, Powai, Mumbai- 400076,
India}

\begin{abstract}
We have studied the electronic structure of \( 3d \) transition-metal-\( MgB_{2} \)
alloys, \( Mg_{0.97}TM_{0.03}B_{2} \), \( (TM\equiv Sc,\, Ti,\, V,\, Cr\, Mn,\, Fe,\, Co,\, Ni,\, Cu,\, Zn) \)
using \emph{}Korringa-Kohn-Rostoker coherent-potential approximation (KKR-CPA)
method in the atomic-sphere approximation (ASA). For \emph{unpolarized} calculations,
our results for \( Mg_{0.97}TM_{0.03}B_{2} \) alloys are similar to that of
\( 3d \) impurities in other \( s \) and \( s-p \) metals. In particular,
the local densities of states (DOS) associated with the \( 3d \) impurities
are similar to our earlier work on \( 3d \) impurities in bulk \( Al \) {[}P.
P. Singh, Phys. Rev. B \textbf{43}, 3975 (1991), P. P. Singh, J. Phys.: Condens.
Matter \textbf{3}, 3285 (1991){]}. For \emph{spin-polarized} calculations, we
find only the alloys of \( V,\, Cr,\, Mn,\, Fe \) and \( Co \) with \( MgB_{2} \)
to be magnetic of all the \( 3d \) elements. We also find that \( Cr \) and
\( Mn \) in \( MgB_{2} \) have a relatively large local magnetic moment of
\( 2.43\, \mu _{B} \) and \( 2.87\, \mu _{B} \), respectively. We have used
the \emph{unpolarized}, self-consistent potentials of \( Mg_{0.97}TM_{0.03}B_{2} \)
alloys, obtained within the coherent-potential approximation, to calculate the
electron-phonon coupling constant \( \lambda  \) using the Gaspari-Georffy
formalism and the superconducting transition temperature \( T_{c} \) using
the Allen-Dynes equation. We find that the calculated \( T_{c} \) is the lowest
for \( Mg_{0.97}Cr_{0.03}B_{2} \) and the highest for \( Mg_{0.97}Zn_{0.03}B_{2} \),
in qualitative agreement with experiment. The calculated trend in variation
of \( T_{c} \) from \( Mn \) to \( Zn \) is also similar to the available
experimental data. Our analysis of the variation in \( T_{c} \), in terms of
the DOS and the spectral function along \( \Gamma  \)to \( A \) direction,
shows the variation to be an interplay between the total DOS at the Fermi energy
and the creation/removal of states along \( \Gamma  \) to  \( A \) direction
{[}P. P. Singh, cond-mat/0201093{]}. 
\end{abstract}
\maketitle

\section*{I. Introduction}

The nature of interaction responsible for superconductivity in \( MgB_{2} \)
\cite{nag,bud,hin,tak,rev,yil,kor,an1,kon,boh,pps1,med,sat,bel,liu,cho} suggests
a gradual decrease in the superconducting transition temperature, \( T_{c} \),
upon addition of impurities \cite{xia,zha,kaz,mor} with increasing \emph{electron/atom}
ratio. A systematic increase in the number of available electrons is expected
to fill up the \( \sigma  \) holes in \( MgB_{2} \), which are coupled strongly
to the in-plane bond-stretching mode of \( B \), and thereby reduce the strength
of the electron-phonon coupling resulting in a decrease in \( T_{c} \). The
observed variation in \( T_{c} \) of alloys of \( Al \) with \( MgB_{2} \)
and its understanding in terms of a gradual filling up of \( \sigma  \) holes
provides a good example of this picture \cite{pps2,bar}.

Since the \emph{electron/atom} ratio increases as one goes from \( Sc \) to
\( Zn, \) based upon the above argument one expects the \( T_{c} \) to systematically
decrease as the different elements from the \( 3d \) row are added to \( MgB_{2}. \)
However, the experimentally observed changes in \( T_{c} \) of \( 3d \) transition-metal-\( MgB_{2} \)
alloys \cite{mor} do not seem to follow the expected trend of a systematic
decrease in \( T_{c} \) as one goes from \( Sc \) to \( Zn. \) For example,
a \( 3\% \) \( Mn \)-doped \( MgB_{2} \) \( (Mg_{0.97}Mn_{0.03}B_{2}) \)
has a \( T_{c} \) of only \( 33.1\, K, \) while \( MgB_{2} \) doped similarly
with \( Fe,\, Co \) and \( Ni \) show \( T_{c} \)'s of \( 37.8,\, 35.7 \)
and \( 37.8\, K, \) respectively \cite{mor}. Surprisingly, \( Zn- \)doped
\( MgB_{2} \) shows the highest \( T_{c} \) \( (=38.4\, K) \) of all the
transition-metal-\( MgB_{2} \) alloys investigated so far \cite{kaz,mor}.
The observed changes in \( T_{c} \) of transition-metal-doped \( MgB_{2} \)
alloys, especially as one goes from \( Fe \) to \( Zn, \) are unexpected even
after making allowances for magnetic effects \cite{all1,all2}.

To understand the changes in the electronic structure and the superconducting
properties of \( MgB_{2} \) alloys upon addition of \( 3d \) transition-metal
impurities, we have carried out \( ab\, initio \) studies of \( Mg_{0.97}TM_{0.03}B_{2} \)
\( (TM\equiv Sc,\, Ti,\, V,\, Cr\, Mn,\, Fe,\, Co,\, Ni,\, Cu,\, Zn) \) alloys
using density-functional-based methods. We have used Korringa-Kohn-Rostoker
coherent-potential approximation method \cite{fau} in the atomic-sphere approximation
(KKR-ASA CPA) \cite{pps_cpa} for taking into account the effects of disorder,
Gaspari-Gyorffy formalism \cite{gas} for calculating the electron-phonon coupling
constant \( \lambda  \), and Allen-Dynes equation \cite{all1,all2} for calculating
\( T_{c} \) in \( Mg_{0.97}TM_{0.03}B_{2} \) alloys. For understanding the
variation in \( T_{c} \) as one goes from \( Sc \) to \( Zn \), we have analyzed
our results in terms of the changes in the spectral function \cite{fau} along
\( \Gamma  \) to \( A \) \cite{pps2,pps_high} and the densities of states
(DOS), in particular, the changes in the \( B \) \( p \) and the transition-metal
\( d \) contributions to the total DOS. The changes in the magnetic alloys
are described in terms of spin-resolved densities of states and the local magnetic
moments. Before we describe our results, we outline some of the computational
details of our calculation.

\section*{II. Computational Details}

The charge self-consistent electronic structure of unpolarized as well as spin-polarized
\( Mg_{0.97}TM_{0.03}B_{2} \) alloys has been calculated using the KKR-ASA
CPA method. We have used the CPA successfully to describe the electronic structure
of \( Al- \)doped \( MgB_{2} \) alloys \cite{pps2}. We parametrized the exchange-correlation
potential as suggested by Perdew-Wang within the generalized gradient approximation
\cite{perdew}. The Brillouin zone (BZ) integration was carried out using \( 1215 \)
\textbf{k}-points in the irreducible part of the BZ. For DOS calculations, we
added a small imaginary component of \( 1 \) \( mRy \) to the energy and used
\( 4900 \) \textbf{k}-points in the irreducible part of the BZ. The lattice
constants for \( Mg_{0.97}TM_{0.03}B_{2} \) were fixed at the \( MgB_{2} \)
values. The Wigner-Seitz radius for \( Mg \) was slightly larger than that
of \( B \), while the Wigner-Seitz radii of the impurities were equal to their
bulk value as given in Ref. \cite{skriver}. The maximum \( l \) used was \( l_{max} \)
= \( 3 \).

As indicated above, the electron-phonon coupling constant \( \lambda  \) was
calculated using Gaspari-Gyorffy formalism with the charge self-consistent potentials
of unpolarized \( Mg_{0.97}TM_{0.03}B_{2} \) alloys obtained with the KKR-ASA
CPA method. Subsequently, the variation of \( T_{c} \) was calculated using
Allen-Dynes equation. The average value of phonon frequency\( \omega _{ln} \)
for \( MgB_{2} \) was taken from Ref. \cite{kon} and \( \mu ^{*}=0.09 \).

\section*{III. Results and Discussion}

In this section we present the results of our self-consistent electronic structure
calculations for \( Mg_{0.97}TM_{0.03}B_{2} \) alloys. We first describe the
results of the \emph{unpolarized} calculations in terms of the total and the
sub-lattice resolved DOS, including the local DOS due to the \( 3d \) transition-metal
impurity. From our \emph{spin-polarized} calculations, described next, we find
only the alloys of \( V,\, Cr,\, Mn,\, Fe \) and \( Co \) with \( MgB_{2} \)
to be magnetic of all the \( 3d \) elements. These results are discussed using
the spin-resolved DOS and the local moments. Finally, the calculated variation
in \( T_{c} \) for \( Mg_{0.97}TM_{0.03}B_{2} \) alloys is compared with the
available experimental data, and analyzed in terms of the DOS and the spectral
function along \( \Gamma  \) to \( A \).

\subsection*{\emph{A. Unpolarized Total and Local Densities of States}}

In Fig. 1 we show the calculated densities of states of \( Mg_{0.97}TM_{0.03}B_{2} \)
alloys from \( Sc \) to \( Zn \), including the total DOS, \( N(E), \) and
the concentration-weighted, sub-lattice resolved DOS for the two inequivalent
sub-lattices \( (N^{sub}(E)) \) in \( MgB_{2} \). The two inequivalent sub-lattices
consist of the \( Mg \) and the transition-metal on one sub-lattice \( (N^{sub}_{1}(E)=0.97N^{sub}_{Mg}(E)+0.03N^{sub}_{TM}(E)) \)
and \( B \) on the other two sub-lattices \( (N^{sub}_{2}(E)=2N^{sub}_{B}(E)) \).
As we go from \( Sc \) to \( Zn \), the increase in the \emph{electron/atom}
ratio moves the Fermi energy, \( E_{F} \), as well as increases the hybridization
of the host \( s-p \) bands with the impurity \( d- \)band. The change in
\( E_{F} \) has important consequences for the superconducting properties of
these alloys, as discussed later. The inward movement of the \( d \)-band as
we go from \( Sc \) to \( Zn \) is clearly manifested in the total DOS as
well as the concentration-weighted DOS of \( Mg \) sub-lattice. However, we
find that the DOS at the \( B \) sub-lattice remains largely unaffected due
to the transition-metal impurity, although the movement of \( E_{F} \) changes
the \( B \) contribution to the total DOS at \( E_{F} \). 

\begin{figure*}
{\par\centering \resizebox*{!}{8.6cm}{\rotatebox{-90}{\includegraphics{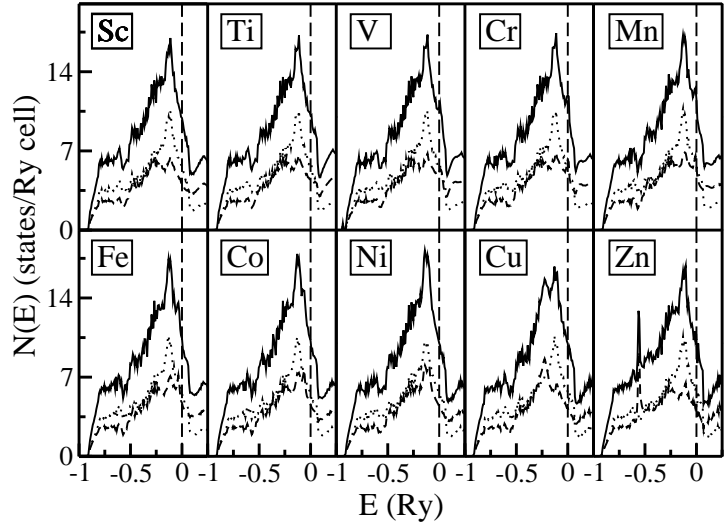}}} \par}

\caption{The calculated total density of states (solid line) for \protect\( Mg_{0.97}TM_{0.03}B_{2}\protect \)
alloys. The contributions to the total DOS from the \protect\( Mg\protect \)
sub-lattice (dashed line) and the \protect\( B\protect \) sub-lattice (dotted
line) are also shown. Note that the \protect\( Mg\protect \) sub-lattice contains
both \protect\( Mg\protect \) and the transition-metal \protect\( (TM)\protect \)
impurity atoms. The vertical dashed line indicates the Fermi energy.}
\end{figure*}

\begin{figure*}
{\par\centering \resizebox*{!}{8.6cm}{\rotatebox{-90}{\includegraphics{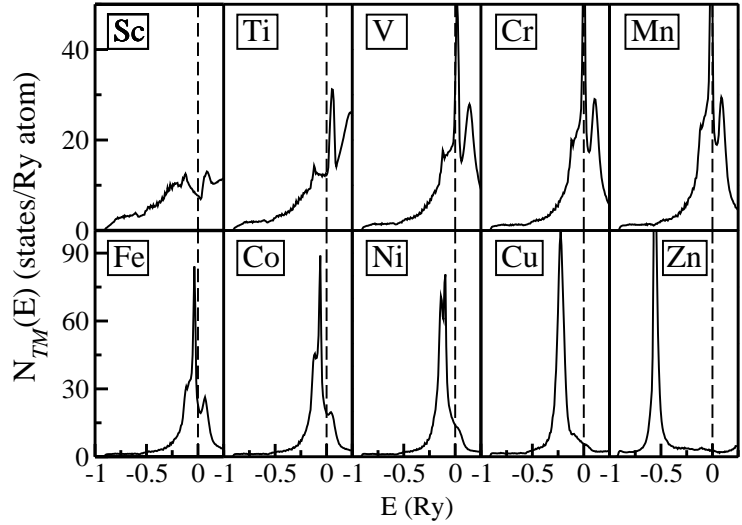}}} \par}

\caption{The calculated total local density of states (solid line) for transition-metal
impurity in \protect\( Mg_{0.97}TM_{0.03}B_{2}\protect \) alloys. The vertical
dashed line indicates the Fermi energy.}
\end{figure*}

To be able to examine the changes in the DOS of the transition-metal impurities
and to show the movement of the \( d- \)band as we go from \( Sc \) to \( Zn \),
we show in Fig. 2 the DOS of the transition metal impurity \( N^{sub}_{TM}(E) \)
on the \( Mg \) sub-lattice in \( Mg_{0.97}TM_{0.03}B_{2} \) alloys. We find
that the \( d- \)level crosses \( E_{F} \) between \( Cr \) and \( Mn \),
resulting in a very high density of states at \( E_{F} \) for these alloys.
The relatively high DOS for \( N^{sub}_{Cr}(E_{F}) \) and \( N^{sub}_{Mn}(E_{F}) \)
points to the possibility of local magnetic moment formation at the impurity
sites. By the time we come to \( Zn \) impurity, the \( d- \)level is well
inside \( E_{F} \) and the impurity contribution to the total DOS has reduced
significantly. We also find that the lowering of \( d- \)level with increasing
\( d- \)electrons is accompanied by the narrowing of the \( d- \)band. For
example, in the case of \( Zn \) impurity the width of the \( d- \)band is
the narrowest, indicating almost an atomic-like behavior of these \( d- \)electrons.
It is interesting to note that these results are qualitatively similar to our
results on \( 3d \) transition-metal impurities in \( Al \) \cite{pps_rec,pps_dil}

\subsection*{B. Spin-Polarized Total and Local Densities of States}

The results of our spin-polarized calculations for \( Mg_{0.97}TM_{0.03}B_{2} \)
alloys are shown in Figs. 3-5. Our calculations show that the alloys of \( V,\, Cr,\, Mn,\, Fe \)
and \( Co \) in \( MgB_{2} \) are exchange-split, as can be seen in Fig. 3
where we have plotted the spin-polarized total and the concentration-weighted,
sub-lattice resolved DOS for \( V,\, Cr,\, Mn,\, Fe \) and \( Co \) in \( MgB_{2} \).
Once again, we find that the \( B \) sub-lattice remains unaffected due to
magnetic moment formation at the impurity site on the \( Mg \) sub-lattice.
To clearly show the changes in the majority and the minority spin DOS of the
\( Mg \) sub-lattice due to exchange-splitting in the DOS of the impurity atom,
we show in Fig. 4 the spin-resolved total DOS due to the impurity atom. As expected
from the unpolarized calculations, the exchange splitting is the largest for
\( Cr \) and \( Mn \) alloys. The exchange-splitting in the DOS due to the
impurity atom leads to local moment formation at these sites with \( Cr \)
and \( Mn \) having local moments of \( 2.43\, \mu _{B} \) and \( 2.87\, \mu _{B} \),
respectively. The local moment on \( Co \) is very small (\( 0.01\, \mu _{B} \)).
Note that the total DOS of the impurity is dominated by the \( d \) contribution
and the local moments arise almost entirely due to the impurity \( d \) electrons. 

\begin{figure*}
{\par\centering \resizebox*{!}{8.6cm}{\rotatebox{-90}{\includegraphics{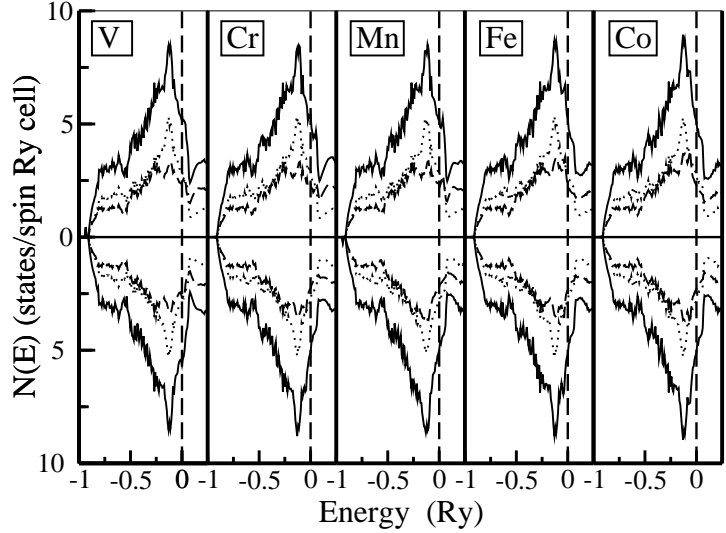}}} \par}

\caption{The calculated total density of states (solid line) for \protect\( Mg_{0.97}TM_{0.03}B_{2}\protect \)
alloys with majority (upper panel) and minority spins (lower panel). The majority
and minority spins contributions to the total DOS from the \protect\( Mg\protect \)
sub-lattice (dashed line) and the \protect\( B\protect \) sub-lattice (dotted
line) are shown in their respective panels. The vertical dashed line indicates
the Fermi energy.}
\end{figure*}

\begin{figure*}
{\par\centering \resizebox*{!}{8.6cm}{\rotatebox{-90}{\includegraphics{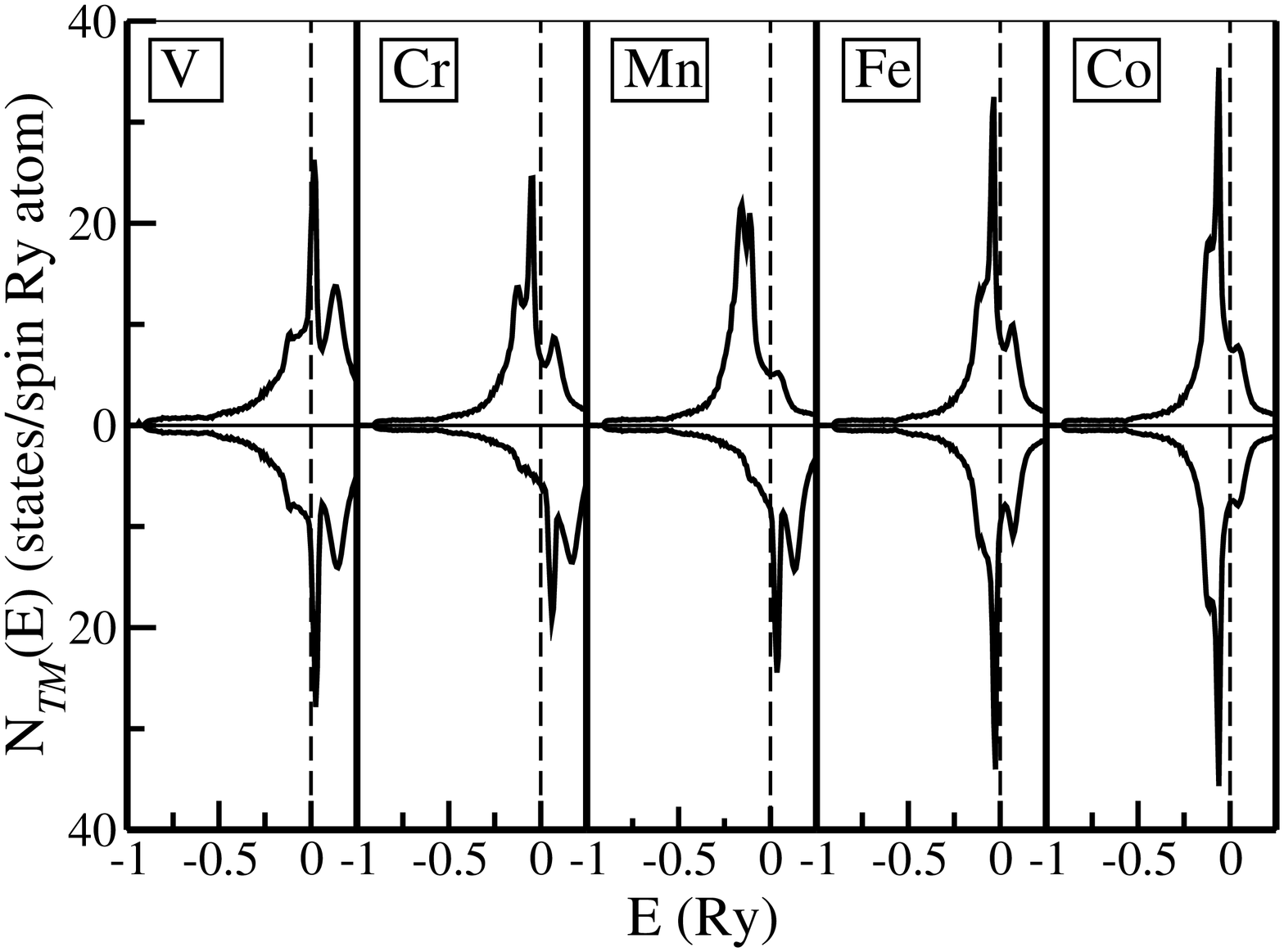}}} \par}

\caption{The calculated total local density of states (solid line) for transition-metal
impurity in \protect\( Mg_{0.97}TM_{0.03}B_{2}\protect \) alloys with majority
(upper panel) and minority spins (lower panel). The vertical dashed line indicates
the Fermi energy.}
\end{figure*}

\begin{figure}
{\par\centering \resizebox*{!}{8.6cm}{\rotatebox{-90}{\includegraphics{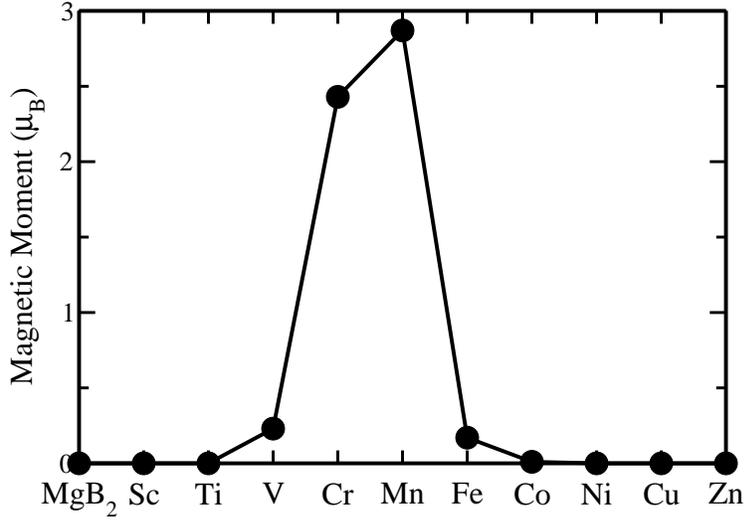}}} \par}

\caption{The calculated local magnetic moment (filled circle) of the transition-metal
impurity in \protect\( Mg_{0.97}TM_{0.03}B_{2}\protect \) alloys.}
\end{figure}

It must be pointed out that the calculation of local moments in alloys is sensitive
to the volume that one associates with the impurity atom. Given the availability
of electrons around \( Mg \) sub-lattice in \( MgB_{2} \) alloys, a judicious
choice of the impurity-atom volume is essential for a reliable description of
the local moments. In our calculations, we have chosen the impurity atomic volume
to be equal to the observed bulk value as given in Ref. {[}\cite{skriver}{]}.
However, it is clear that \( Cr \) and \( Mn \) will show large local moments
in \( MgB_{2} \). 

\begin{figure}
{\par\centering \resizebox*{!}{8.6cm}{\rotatebox{-90}{\includegraphics{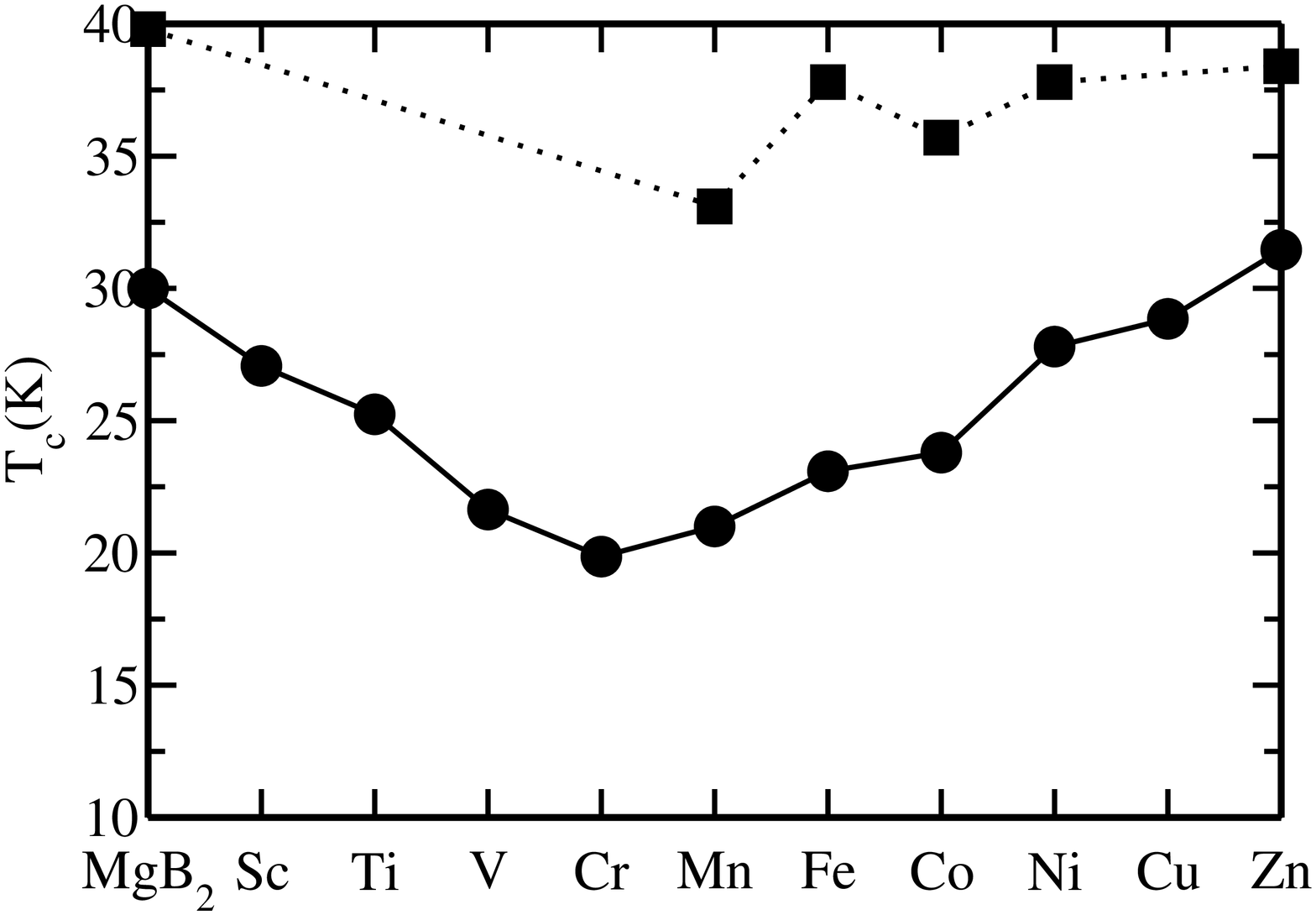}}} \par}

\caption{The calculated (filled circle) and the experimental (filled square) variation
of \protect\( T_{c}\protect \) in \protect\( Mg_{0.97}TM_{0.03}B_{2}\protect \)
alloys.}
\end{figure}

\subsection*{C. Superconducting Transition Temperature}

We have studied the electronic structure of \( Mg_{0.97}TM_{0.03}B_{2} \) alloys
as a prelude to understanding the superconducting properties, especially the
superconducting transition temperature \( T_{c} \), of these alloys. In Fig.
6 we show our calculated \( T_{c} \) as well as the observed \( T_{c} \) \cite{kaz,mor}
for \( Mg_{0.97}TM_{0.03}B_{2} \) alloys. The calculated variation in \( T_{c} \)
across the \( 3d \) row is similar to the one observed experimentally for \( Mn \)
to \( Zn \) in \( MgB_{2} \) \cite{kaz,mor}. Note that our calculated \( T_{c} \)
for \( MgB_{2} \) is equal to \( \sim 30\, K \), which is consistent with
the results of other works \cite{kor,kon,boh} with similar approximations. 

The total DOS and the spectral function along \( \Gamma  \) to \( A \) at
\( E_{F} \) are expected to play an important role in deciding the \( T_{c} \)
of \( Mg_{0.97}TM_{0.03}B_{2} \) alloys. Thus, in the following we examine
the changes in these two quantities in \( Mg_{0.97}TM_{0.03}B_{2} \) alloys
as we go from \( Sc \) to \( Zn \). In Fig. 7 we show the total DOS, \( N(E_{F}) \),
the \( B\, p \) contribution to \( N^{sub}_{B}(E_{F}) \) and the impurity
\( d \) contribution to \( N^{sub}_{TM}(E_{F}) \) at the Fermi energy. Our
calculations show \( Mg_{0.097}Cr_{0.03}B_{2} \) to have the lowest \( T_{c} \)
of all the \( 3d \) alloys, which coincides with the highest \( N(E_{F}) \)
of \( 11.93\, st/(Ry\, cell) \) as well as the highest \( d \) contribution
to the \( N^{sub}_{Cr}(E_{F}) \) of \( 66.15\, st/(Ry\, atom) \), as can be
seen in Fig. 7. In contrast, the \( B\, p \) contribution is not enhanced resulting
in the lowest \( T_{c} \) for \( Mg_{0.97}Cr_{0.03}B_{2} \) within our approach.
Similarly, the \( T_{c} \) for \( Mg_{0.97}Mn_{0.03}B_{2} \) is also small
in comparison with other alloys. Here, we like to point out that the exchange-splitting
will reduce \( N^{sub}_{TM}(E_{F}) \), leading to a smaller \( N(E_{F}) \)
in the case of magnetic alloys. Thus, within our approach, it would have led
to an increase in \( T_{c} \). However, the inclusion of magnetism with its
pair-breaking effects \cite{all1,all2} will further reduce the \( T_{c} \).
The gradual increase in \( T_{c} \) from \( Mn \) to \( Zn \) is due to enhanced
\( B\, p \) contribution to \( N^{sub}_{B}(E_{F}) \) as well as a substantial
decrease in the impurity \( d \) contribution to \( N^{sub}_{TM}(E_{F}) \)
as shown in Fig. 7. 

\begin{figure}
{\par\centering \resizebox*{!}{8.6cm}{\rotatebox{-90}{\includegraphics{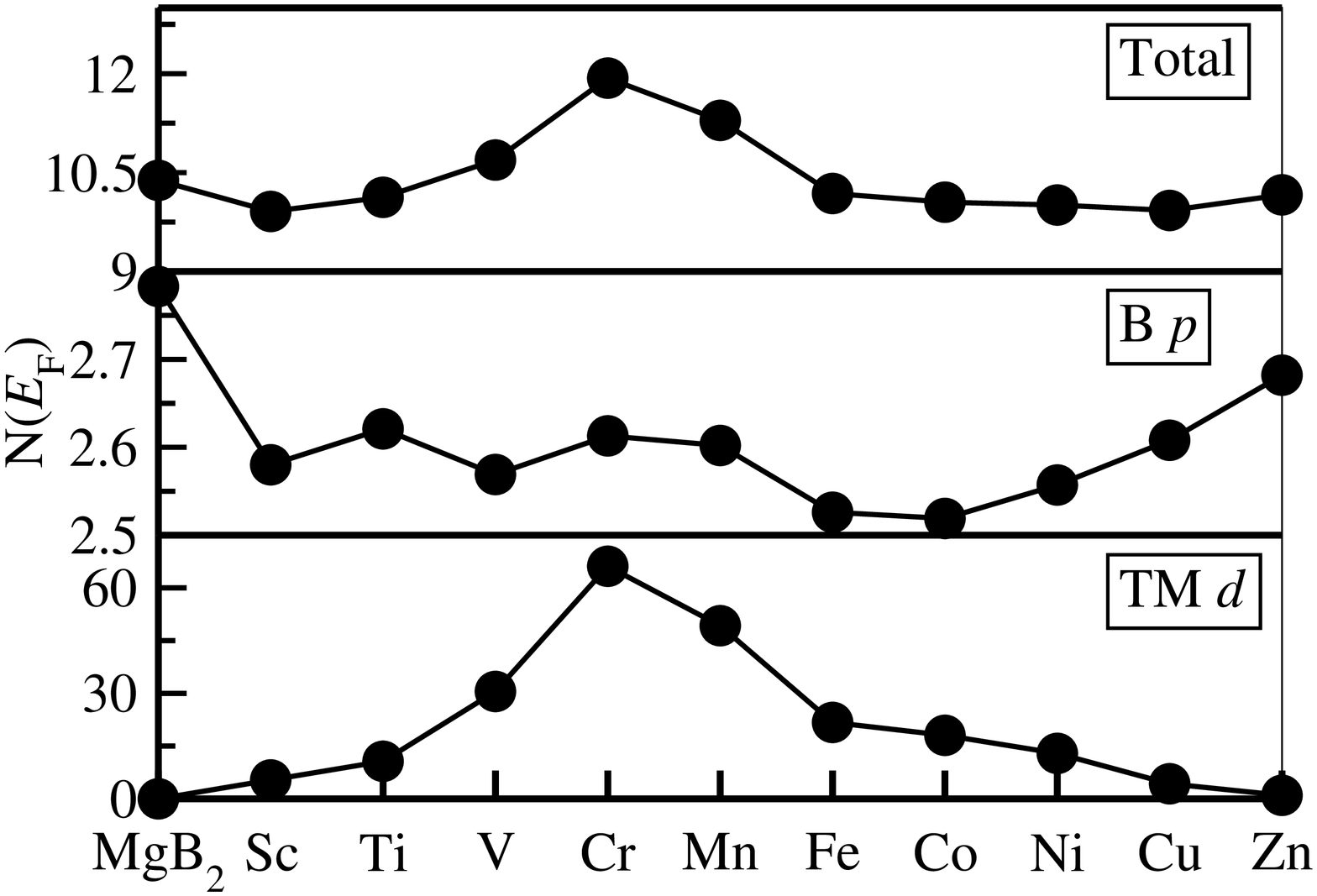}}} \par}

\caption{The calculated total density of states at the Fermi energy (filled circle,
upper panel), \protect\( B\, p\protect \) contribution (filled circle, middle
panel) and the transition-metal \protect\( (TM)\protect \) \protect\( d\protect \)
contribution (filled circle, lower panel) to the total DOS in \protect\( Mg_{0.97}TM_{0.03}B_{2}\protect \)
alloys. }
\end{figure}

To further clarify the reasons for the changes in \( T_{c} \) in \( Mg_{0.97}TM_{0.03}B_{2} \)
alloys, as we go from \( Sc \) to \( Zn \), we show in Fig. 8(a)-(d) the spectral
function, \( A( \)\textbf{k},\( E_{F}) \), calculated along \( \Gamma  \)
to \( A \) direction in \( Mg_{0.97}TM_{0.03}B_{2} \) alloys. For comparison
we have also plotted \( A( \)\textbf{k},\( E_{F}) \) for \( MgB_{2} \) in
Figs. 8(a) and (d). As described in Ref. \cite{pps2}, the additional states
created along \( \Gamma  \) to \( A \) direction reduce the coupling between
the hole-like cylindrical Fermi sheets with the phonons which, in turn, reduces
the \( T_{c} \). Thus an alloy with \( A( \)\textbf{k},\( E_{F}) \) similar
to that of \( MgB_{2} \) will show a \( T_{c} \) close to that of \( MgB_{2} \).
Carrying out such a comparison in Fig. 8, we find that \( Mg_{0.97}Zn_{0.03}B_{2} \)
will have a \( T_{c} \) close to that of \( MgB_{2} \), while \( Mg_{0.97}V_{0.03}B_{2} \)
will show the lowest \( T_{c} \) of them all. The fact that our calculation
shows \( Mg_{0.97}Cr_{0.03}B_{2} \) to have a lower \( T_{c} \) than \( Mg_{0.97}V_{0.03}B_{2} \)
is due to the relatively high impurity DOS at \( E_{F} \). In addition, in
the case of \( Cr \) and \( Mn \) impurity in \( MgB_{2} \), the magnetic
effects will further reduce the \( T_{c} \) \cite{all2}. These results reinforce
the observation made in Ref. \cite{pps2} that the way \( T_{c} \) changes
in \( MgB_{2} \) upon alloying depends dramatically on the location of the
added/modified \textbf{k}-resolved states on the Fermi surface.
\begin{figure}
{\par\centering \resizebox*{!}{8.6cm}{\rotatebox{-90}{\includegraphics{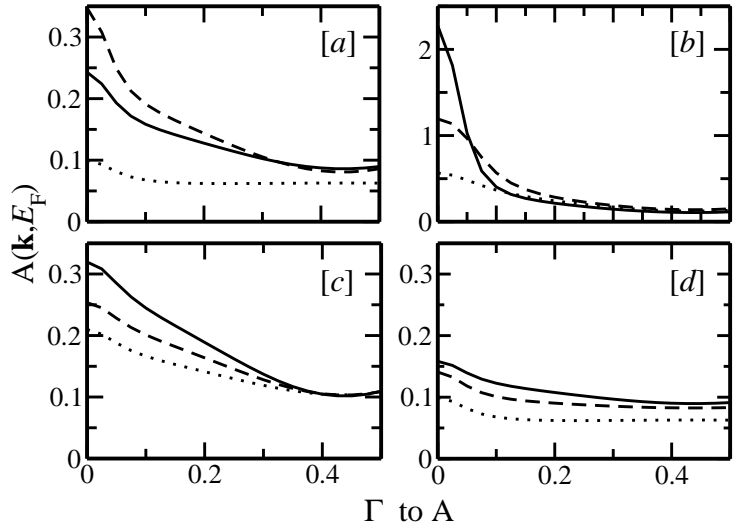}}} \par}

\caption{The calculated spectral function along \protect\( \Gamma \protect \) to \protect\( A\protect \),
evaluated at the Fermi energy in \protect\( Mg_{0.97}TM_{0.03}B_{2}\protect \)
alloys. In figures \protect\( (a)\protect \)-\protect\( (d)\protect \) the
solid, dashed and dotted lines correspond to \protect\( (a)\protect \) \protect\( Sc,\protect \)
\protect\( Ti\protect \) and \protect\( MgB_{2}\protect \), \protect\( (b)\protect \)
\protect\( V,\protect \) \protect\( Cr\protect \) and \protect\( Mn\protect \),
\protect\( (c)\protect \) \protect\( Fe,\protect \) \protect\( Co\protect \)
and \protect\( Ni\protect \), and \protect\( (d)\protect \) \protect\( Cu,\protect \)
\protect\( Zn\protect \) and \protect\( MgB_{2}\protect \) alloys, respectively.
Note the change in the vertical scale in \protect\( (b)\protect \).}
\end{figure}

\section*{IV. Summary}

We have studied the electronic structure of \( 3d \) transition-metal-\( MgB_{2} \)
alloys using \emph{}Korringa-Kohn-Rostoker coherent-potential approximation
method in the atomic-sphere approximation. Our results for the \emph{unpolarized}
calculations are similar to that of \( 3d \) impurities in other \( s \) and
\( s-p \) metals. From \emph{spin-polarized} calculations we find that only
alloys of \( V,\, Cr,\, Mn,\, Fe \) and \( Co \) in \( MgB_{2} \) are magnetic,
with \( Cr \) and \( Mn \) having the largest local magnetic moment of \( 2.43\, \mu _{B} \)
and \( 2.87\, \mu _{B} \), respectively. We have used the \emph{unpolarized},
self-consistent potentials of \( Mg_{0.97}TM_{0.03}B_{2} \) alloys, obtained
within the coherent-potential approximation, to calculate the electron-phonon
coupling constant \( \lambda  \) using the Gaspari-Georffy formalism. Then,
with the help of Allen-Dynes equation we have calculated the superconducting
transition temperature \( T_{c} \) of these alloys. We find that the calculated
\( T_{c} \) is the lowest for \( Mg_{0.97}Cr_{0.03}B_{2} \) and the highest
for \( Mg_{0.97}Zn_{0.03}B_{2} \), in qualitative agreement with experiment.
The trend in variation of \( T_{c} \) from \( Mn \) to \( Zn \) is also similar
to the available experimental data.

\end{document}